\documentclass[12pt,preprint]{emulateapj}
\usepackage{natbib}

\shorttitle{SONYC - Chamaeleon I}
\shortauthors{Mu\v{z}i\'c et al.}

\begin{document}
\bibliographystyle{apj}


\title{Substellar Objects in Nearby Young Clusters (SONYC) III: Chamaeleon-I \footnote{Based on observations collected at the European Organisation for Astronomical Research in the Southern Hemisphere, Chile (programs 078.C-0049 and 382.C-0174)}}


\author{Koraljka Mu\v{z}i\'c\altaffilmark{1}, Alexander Scholz\altaffilmark{2},
Vincent Geers\altaffilmark{3}, Laura Fissel\altaffilmark{1},
Ray Jayawardhana\altaffilmark{1,**}}

\email{muzic@astro.utoronto.ca}

\altaffiltext{1}{Department of Astronomy \& Astrophysics, University of Toronto, 50 St. George Street, Toronto, ON M5S 3H4, Canada}
\altaffiltext{2}{School of Cosmic Physics, Dublin Institute for Advanced Studies, 31 Fitzwilliam Place, Dublin 2, Ireland}
\altaffiltext{3}{Institut f\"ur Astronomie, ETH, Wolfgang-Pauli-Strasse 27, 8093 Z\"urich, Switzerland}
\altaffiltext{**}{Principal Investigator of SONYC}

\begin{abstract}
SONYC -- {\it Substellar Objects in Nearby Young Clusters} -- is a survey program to investigate the frequency and properties of substellar objects with masses down to a few times that of Jupiter in nearby star-forming regions.
In this third paper, we present our recent results in the Chamaeleon-I star forming region. We have carried out
deep optical and near-infrared imaging in four bands ($I, z, J, K_S$) using VIMOS on the ESO Very Large Telescope
and SOFI on the New Technology Telescope, and combined
our data with mid-infrared data from the Spitzer Space Telescope. The survey covers 
$\sim\,$0.25$\,$deg$^2$ on the sky, and reaches completeness limits of 23.0 in the I-band, 18.3 in the J-band, and 
16.7 in $K_S$-band.
Follow-up spectroscopy of the candidates selected from the optical photometry ($I\lesssim 21$) was carried out using the multi-object spectrograph VIMOS on the VLT. We identify 13 objects consistent with M spectral types, 11 of
which are previously known M-dwarfs with confirmed membership in the cluster. The 2 newly reported objects have 
effective temperatures above the substellar limit. We also present two new candidate members of Chamaeleon-I, selected from our $JK$ photometry combined with the Spitzer data. Based on the results of our survey, we estimate that the number of missing very-low-mass members down to $\sim0.008 M_{\odot}$ and $A_V \leq 5$ is $\leq 7$, i.e. $\leq 3 \%$ of the total cluster population according to the current census.
\end{abstract}


\section{Introduction}
\label{s1}
\begin{figure*}
\center
\includegraphics[width=15cm,angle=0]{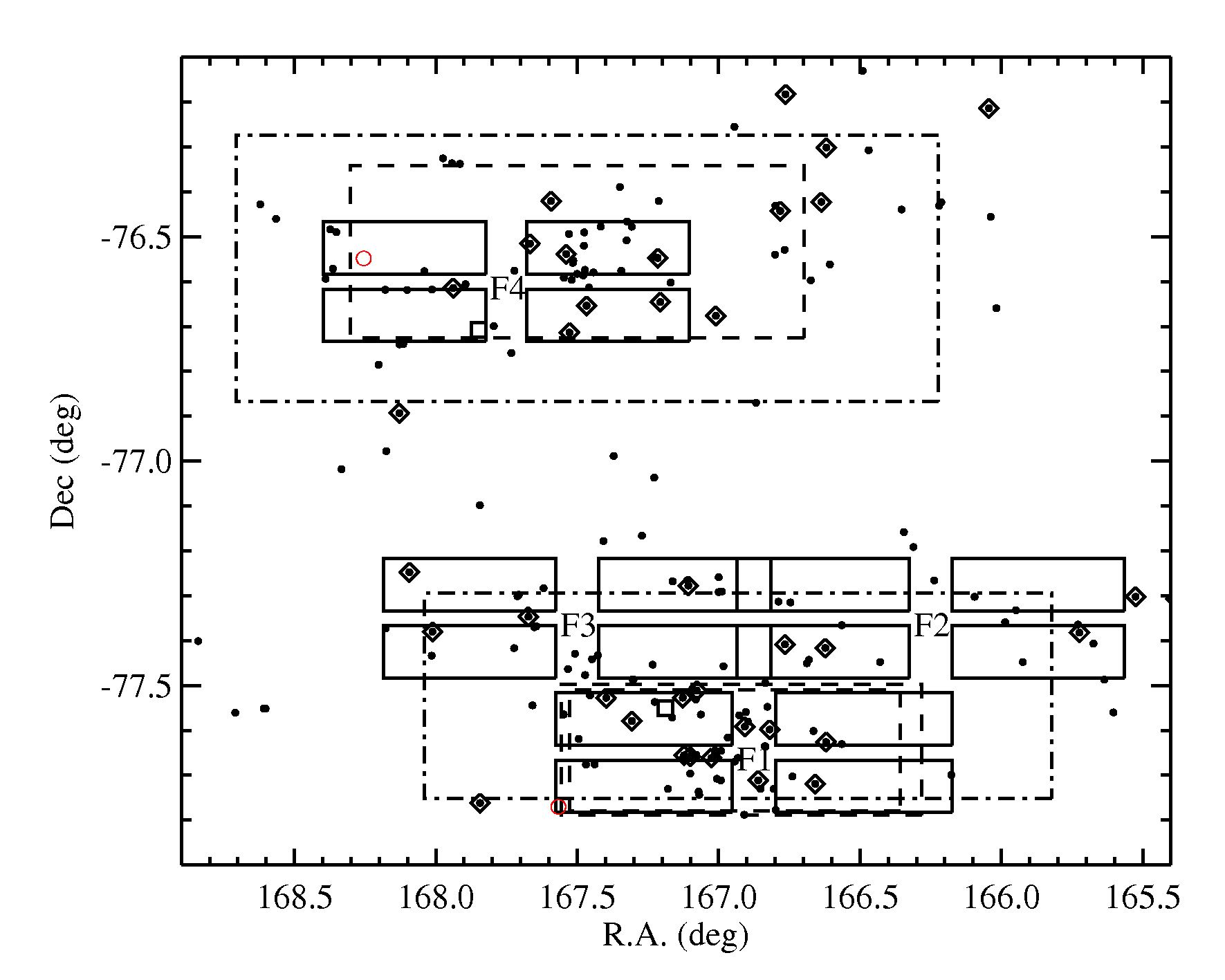}
\caption{Projected distribution of the observed fields in Cha-I. 
Solid lines show four fields observed by VIMOS in the imaging mode. Each field consists
of four quadrants; labeling of the VIMOS fields (F1 to F4) is shown near the centers of 
the gaps between the quadrants. Spatial distribution of the SOFI fields approximately matches the VIMOS fields. 
IRAC fields are outlined with dashed (deep) and dash-dotted line (shallow).
Optical spectra were obtained with VIMOS in the MOS mode, for the objects in fields F1, F3, and F4.
Black dots mark the known members \citep{luhman07, luhman&muench08, luhman08}, with brown dwarfs ($T_{\mathrm{eff}}<3000\,$K) additionally shown as diamonds.
Squares mark the two candidates from Table \ref{tJKSp}, and red crosses mark the two new candidate members from Table \ref{tspec}.}
\label{spatial}
\end{figure*}

Brown dwarfs ($M\lesssim 0.08\,M_{\odot}$) are an important class of object to test the mass dependence in the formation and early evolution of stars and planets. As of today, the origin of these substellar objects it not clear \citep{whitworth05}; they may be the result of turbulent fragmentation \citep{padoan&nordlund04} or ejections from multiple systems \citep{bate09a}. In addition, brown dwarfs might have been ejected from protoplanetary disks \citep{stamatellos09} where they have formed by disk fragmentation, and we might observe the overlap of star and planet forming processes. Further issues that can be addressed by investigating young brown dwarfs are, among others, the evolution of accretion disks (e.g., the impact of photoevaporation), the formation of multiple systems \citep{bate09a}, the mass dependence of planet formation 
\citep{payne&lodato07}, and the early regulation of angular momentum \citep{scholz04,scholz05}.

The prerequisite for tackling these issues is to obtain a reliable picture of the frequency and properties of young brown dwarfs. The currently available surveys are affected by two major biases:
a) In most regions they do not extend below the deuterium-burning limit ($\sim 0.015\,M_{\odot}$), i.e. they do not cover the regime of free-floating objects with planetary masses. In several regions, however, it has been shown that the mass function extends to 0.01$\,M_{\odot}$ \citep{zapateroosorio00, lucas05, luhman08} or even below, and a lower mass limit (the ``bottom of the IMF'') has not been found yet, which indicates that our census of substellar objects is incomplete at the lowest masses; b)~Some of the deepest surveys are biased by design, either because they are based on mid-infrared data from Spitzer and thus will only find objects surrounded by dusty disks (e.g., \citealt{luhman08}) or because they are carried out using narrow-band methane imaging, which probes only a specific range of spectral types (e.g., \citealt{haisch10}). While samples obtained with these methods are useful to constrain the mass function, they are of limited use for other purposes (for example, Spitzer samples cannot reliably probe disk lifetimes as a function of mass).

Our program SONYC (Substellar Objects in Nearby Young Clusters) aims to mitigate these selection effects and to help in establishing a more complete census of brown dwarfs. The project is based on three basic principles: \\
1) By making extensive use of 4- to 8-m-class telescopes, we aim to reach mass limits of $0.005\,M_{\odot}$ and below, with the specific goal of probing the bottom of the IMF. \\
2) The primary means of identifying candidates is broadband imaging in the optical and near-infrared, thus aiming to detect the photosphere. This results in large samples of candidates, i.e. requires extensive follow-up spectroscopy. This dataset is complemented by publicly available Spitzer data. \\
3) By probing several star forming regions we want to probe for environmental differences in the frequency and properties of substellar objects.

So far, we have published results for two regions, NGC$\,$1333 \citep{scholz09a} and $\rho$ Ophiuchus \citep{geers10}. In NGC$\,$1333 we identify 19 substellar objects, 12 of them previously unknown. We tentatively find a cutoff in the mass function between 0.01 and 0.02$\,M_{\odot}$. This may be the first sign that we have reached the bottom of the IMF. In $\rho$ Ophiuchus we spectroscopically confirm a previously unknown brown dwarf, and identify a number of new candidates based on Spitzer data. In this region, however, the current census is highly incomplete due to strong and variable extinction.

In this paper we present the SONYC results for the star forming region Chamaeleon I (hereafter Cha-I). Cha-I harbors a rich population of known YSOs \citep{luhman07}, most of them grouped in two clusters, each dominated by a bright A star. Cha-I is less compact than NGC1333 and has less extinction than $\rho$-Oph. Previous surveys have identified a significant number of substellar objects \citep{comeron99,neuhauser&comeron99,oasa99,comeron04, lopezmarti04}. Extensive photometric and spectroscopic studies of the region 
\citep{luhman04, luhman07,luhman&muench08,luhman08} result in 40 spectroscopically confirmed objects below the hydrogen-burning limit. Among these are a few objects with estimated masses at or below the deuterium-burning limit \citep{luhman06, luhman08}. 
Throughout this paper we will use the current membership census, referred to as the ``Luhman census'', by Luhman (2007, 215 objects, his table 6), complemented by the new members in \citet[8 objects]{luhman&muench08} and \citet[4 objects]{luhman08}. 

This paper is organized as follows. Observations and data reduction are described in 
Section~\ref{s2}. In Section~\ref{s3} we describe selection of substellar candidates from photometry, and the spectroscopic
follow-up of the candidate objects in Cha-I. The results are discussed in Section~\ref{s4}, and conclusions are presented in Section~\ref{s5}.
\begin{figure*}
\center
\includegraphics[width=18cm]{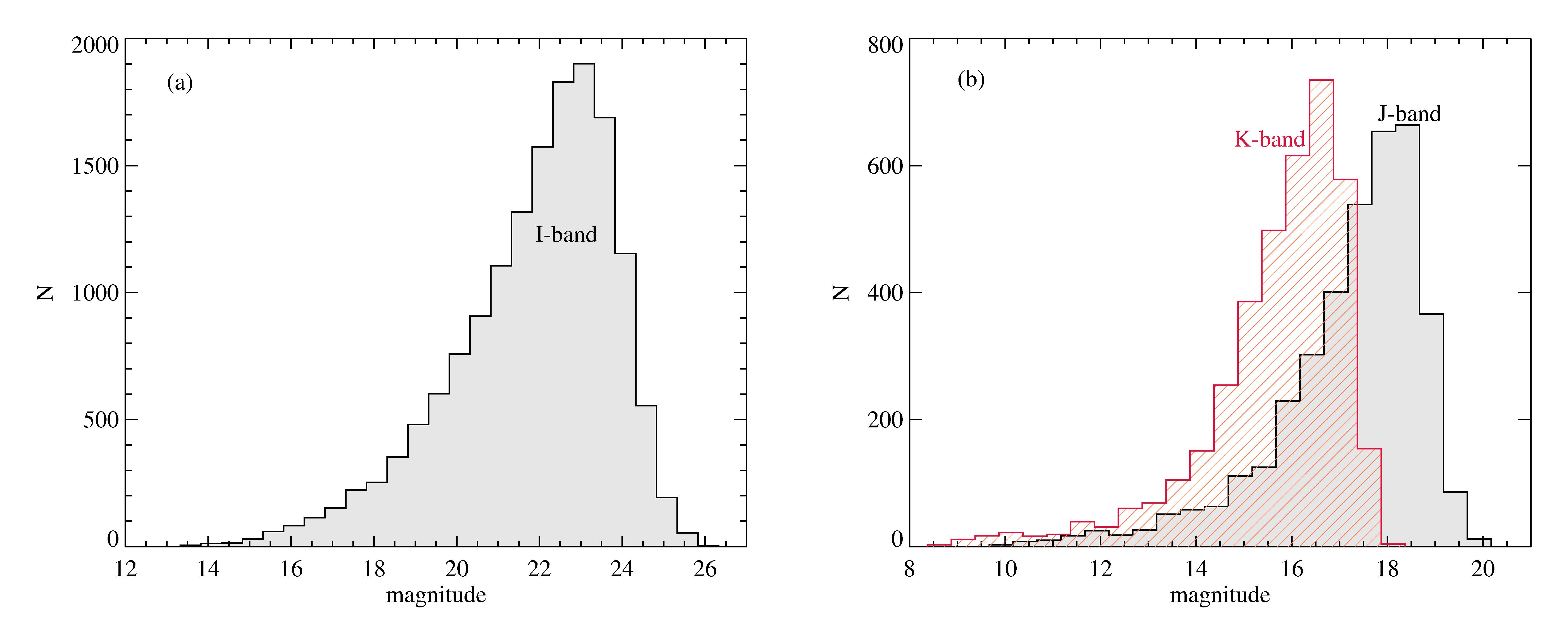}
\caption{Histograms of the objects in our (a) optical and (b) near-infrared photometric catalogs.
The peaks at $I=23.0$, $J=18.3$, and $K_S=16.7$ indicate the completeness limit in each band.
The faintest objects in the survey are found at $I>25$, $J>19.5$, and $K_S>17.5$.}
\label{histogram}
\end{figure*}

\section{Observations and data reduction}
\label{s2}

This study is primarily based on data from the European Southern Observatory, obtained in runs 078.C-0049 and 382.C-0174. The first campaign provided deep optical and near-infrared imaging. From the optical dataset we selected a list of candidate members which were observed using multi-object spectroscopy in the second run. Additionally, we made use of the archival Spitzer images for our target regions. In the following we describe the observations, the basic reduction and photometry.
The fields covered by our surveys are outlined in Figure~\ref{spatial}.

\subsection{Optical imaging}
\label{s21}

We used VIMOS at the VLT to observe Cha-I in the optical wavelength regime. 
VIMOS is a wide-field imager with 4 CCDs arranged in a $2\times 2$ array. Each detector covers  a field of view of $7'\times8'$ with a pixel resolution of 0\farcs2. The four quadrants are separated by gaps of approximately $2'$. Four pointings were observed, each in the $I$- and $z$-band filters. 
See Table~\ref{tfields} for more details about the VIMOS imaging run.
With this observing strategy we do not cover the gaps between the chips, i.e. the survey is spatially incomplete.  Most of the currently known young population of Cha-I is clustered around the two bright stars HD$\,$97048 and HD$\,$97300. These two areas (with the bright stars in the central gaps of the VIMOS array) are covered in our fields 1 and 4. The two remaining fields are located adjacent to each other and north of field 1. Taken together, the SONYC fields cover 38\% of the known Cha-I population, including 17 known brown dwarfs.

\begin{deluxetable*}{cllclccl}
\tabletypesize{\scriptsize}
\tablecaption{Observing log for our imaging and spectroscopic surveys.
\label{tfields}}
\tablewidth{0pt}
\tablehead{
\colhead{field} & \colhead{$\alpha$(J2000)} & \colhead{$\delta$(J2000)} & \colhead{filter} &  \colhead{UT date}  &  \colhead{airmass} & \colhead{seeing\tablenotemark{a}} & \colhead{notes} }
\tablecolumns{8}
\startdata
\multicolumn{8}{c}{Optical and near-infrared imaging}\\
\\
1 & 11:07:30 & -77:39:00 & $I$, $z$ & Jan 17, 2007 &  1.7 & 0.9$''$ & $11 \times 150\,$s in $z$, $10 \times 150\,$s in $I$\\
2 & 11:05:00 & -77:21:00 & $I$, $z$ & Feb 10, 2007 & 1.7 & 0.7$''$ & $11 \times 150\,$s in $z$, $10 \times 150\,$s in $I$\\
3 & 11:10:00 & -77:21:00 & $I$, $z$ & Feb 11, 2007 & 1.7 & 0.8$''$ & $10$ (9 for Q3)$ \times 150\,$s in $z$, $11$ (10 for Q4) $\times 150\,$s in $I$\\
4 & 11:11:00 & -76:36:00 & $I$, $z$ & Feb 05, 2007 & 1.6 & 0.5$''$ & $10 \times 150\,$s in $z$, $10 \times 150\,$s in $I$\\
1 & 11:07:30 & -77:39:00 & $J$, $K_S$ & Nov 11, 2006 & 1.9 & 1.0$''$ & $64 \times 20\,$s in $J$, $64 \times 8\,$s in $K_S$\\
2 & 11:05:00 & -77:21:00 & $J$, $K_S$ & Dec 05, 2006 & 2.0 & 1.2$''$ & $64 \times 20\,$s in $J$, $64 \times 8\,$s in $K_S$ \\
3 & 11:10:00 & -77:21:00 & $J$, $K_S$ & Dec 20, 2006 & 1.6 & 0.9$''$ & $64 \times 20\,$s in $J$, $64 \times 8\,$s in $K_S$\\
4 & 11:11:00 & -76:36:00 & $J$, $K_S$ & Nov 16, 2006 & 1.9 & 1.1$''$ & $64 \times 20\,$s in $J$, $64 \times 8\,$s in $K_S$\\
\\
\multicolumn{8}{c}{Spectroscopy}\\
\\
1 & 11:07:30 & -77:39:00 & LR\_red & Feb 16, 2009 & 1.7 & \nodata & $4\times1610$$\,$s \\
3 & 11:10:00 & -77:21:00 & LR\_red & Mar 05, 2009 & 1.7 & \nodata & $4\times1610$$\,$s \\
4 & 11:11:00 & -76:36:00 & LR\_red & Mar 06, 2009 & 1.7 & \nodata & $4\times1610$$\,$s \\
\enddata
\tablenotetext{a}{measured as a typical FWHM of unsaturated sources in our images}
\end{deluxetable*}

A standard image reduction was preformed for each individual science observation, using
the VIMOS pipeline recipes. 
The individual images were then
combined and the world coordinate system was calibrated against objects
found in the 2MASS point source catalog with $17>J>12$.

Sources were identified and extracted from each CCD chip using the
Source Extractor ({\em SExtractor}) software package \citep{SExtractor96}.
For object identification we required at least 5 pixel to be above the
$3\sigma$ threshold detection limit.  {\em SExtractor} automatic
aperture fitting photometry routines were used to calculate the flux
of each detected source.  We further rejected objects with flux signal-to-noise of less 
than 5, overly elongated objects ($a/b\,<2.0$),
and objects which could not be found
in both the $I$ and $z$ catalogs.

To calibrate the I-band photometry we compared our instrumental magnitudes with those in Table$\,$1 of \citet{luhman07}, 
for objects with $I>16$. No photometric calibration was available for $z$-band, so in order to compare
color magnitude diagrams for different CCD chips and fields, we calculated the median instrumental $I-z$ of all catalog objects in the range $20.0<I<17.5$, $-1.5<I-z<1.0$. We then subtracted the difference between the median $I-z$ for each chip and the median for F2, CCD chip \#1.

From the histogram of the magnitudes (see Figure~\ref{histogram}), we derive 
the completeness limit of the optical survey at $I=23.0\pm0.2$ mag, and
the limiting magnitude of about 25 mag.

\subsection{Near-infrared imaging}
\label{s22}

The near-infrared observations were designed to provide $J$- and $K_S$-band photometry for the same area covered in the optical survey with VIMOS. We used SOFI at the NTT in its wide-field mode resulting in a field of view of $4.9'\times 4.9'$ and a pixel scale of 0\farcs288. 
The definitions of the SOFI filters are comparable to the 2MASS bands \citep{carpenter01}. 
See Table~\ref{tfields} for more details about the SOFI run.

Standard data reduction including sky subtraction, bad pixel, flat field, and the detector row cross-talk correction, was performed using IDL and DPUSER\footnote{http://www.mpe.mpg.de/\~{}ott/dpuser} (T. Ott). The 32 integrations per VIMOS quadrant were combined into a single mosaic using a simple shift-and-add algorithm. Shifts between the individual exposures were determined by cross-correlation, performed by the $jitter$ routine (part of the ESO $eclipse$ package; \citealt{devillard97}). The images were then shifted and median averaged.

For the source detection we used {\em SExtractor}, requiring at least 5 pixels with flux above the 3.5$\,\sigma$ detection limit. 
Saturated and extended sources were rejected from further consideration.
For the photometric calibration, we identified all 2MASS sources matching the $J$- and $K_S$-band sources in our images with signal-to-noise better than 10. The number of useful calibrators per field ranges from 85 to 164. With these stars we determined the zeropoint ZP1 and ZP2 and a color coefficients C1 and C2 to transfer our magnitudes ($j$, $k$) to the 2MASS system: $J = j + ZP1 + C1(j-k)$ and $K = k + ZP2 + C2(j-k)$. With one exception, the color terms are 0.02 or smaller, and do not have significant effects on the calibration, which is in agreement with the findings in Carpenter (2001). The zeropoints are 1.15 to 1.23 mag in $J$ and -0.53 to -0.59 mag in $K_S$. The spread can be explained with the variations in airmass between our fields. The typical error in the photometric calibration is 0.06\,mag.
From the histogram of the magnitudes (see Figure~\ref{histogram}), we derive 
the completeness limits of the NIR part of the survey: $J=18.3\pm0.2$ mag, and $K_S=16.7\pm0.2$ mag.
The limiting magnitudes are about $J=19$ and $K_S=17.5$.

The WCS system for all images was determined with {\it ccmap} in IRAF, based on the 2MASS sources. Typically, we used 50-150 2MASS stars per VIMOS quadrant; the resulting accuracy in RA and DEC is $\sim 0\farcs15$.

\subsection{Spitzer photometry}
\label{s23}

In addition to our optical and near-infrared images, we use IRAC data from the Spitzer Space Telescope to identify candidates. The regions of our optical/near-infrared survey are covered by five AORs: 3955968, 6526208, 12620032 from the program \#36 (PI: Fazio), 3651328 from program \#6 (PI: G. Fazio), and 3960320 from program \#37 (PI: G. Fazio). The three first mentioned AORs are much deeper than the other two (100\,s vs. 12\,s integration time). Our analysis is based on the pipeline reduced PBCD images in the four IRAC channels at 3.6, 4.5, 5.8, and 8.0$\,$\micron\ from these five AORs. Using {\tt daofind} within IRAF we searched for objects with peak fluxes exceeding 5 times the noise in the image. This yields significant contamination, particularly in the bright cloud cores of Cha-I. Therefore we required an object to be detected in all four IRAC images of the same AOR within $\pm 1$ pixel. Note that the PBCD images for these AORs are sampled to 0.6"/pix. For the resulting clean object catalogs we carried out aperture photometry using {\tt daophot}. We chose relatively small apertures of 5 pix and a background annulus of 5-10 pix. We determined aperture corrections using a small sample of isolated bright stars, to shift our photometry to the standard aperture of 10 native pixels, as given in the IRAC handbook. The IRAC photometry catalogs were cross-correlated with the object catalog from our deep $J$- and $K_S$-band images. In this combined near/mid-infrared catalogs we will search for objects with the typical mid-infrared excess as expected for sources with disks.

\subsection{Multi-object spectroscopy}

The optical spectra have been obtained using VIMOS/VLT in the Multi-Object Spectroscopy (MOS) mode. The FOV in the MOS mode is the same as in the imaging mode. We have observed three out of four fields covered by the imaging run (F1, F3, and F4), using the low resolution red grism (LR$\_$red). We covered the wavelength regime between 5500~\AA~and 9500~\AA, with the spectral resolution of R$\,\sim$210, slit width of 1$''$ and a spatial
scale of 0\farcs205. The total on-source time for each field was 4$\times$1610\,s. Data reduction was performed using the VIMOS pipeline provided by ESO. Data reduction steps include bias subtraction, flat-field and bad-pixel correction, wavelength and flux calibration, as well as the final extraction of the spectra. As no jittering offsets have been applied between the individual exposures, fringe correction could not be performed. Our spectra therefore suffer strong fringing effects longwards of $\sim\,$8000\,\AA. In total, we have arranged 308 slits across the three observed fields (126 in F1, 81 in F3, and 101 in F4). After discarding empty slits and very noisy spectra, we are 
left with spectra of 229 objects ready for inspection.  

\section{Selection of substellar candidates in Cha-I}
\label{s3}

\subsection{Optical color-magnitude diagram}
\label{s31}

\begin{figure*}
\center
\includegraphics[width=14cm]{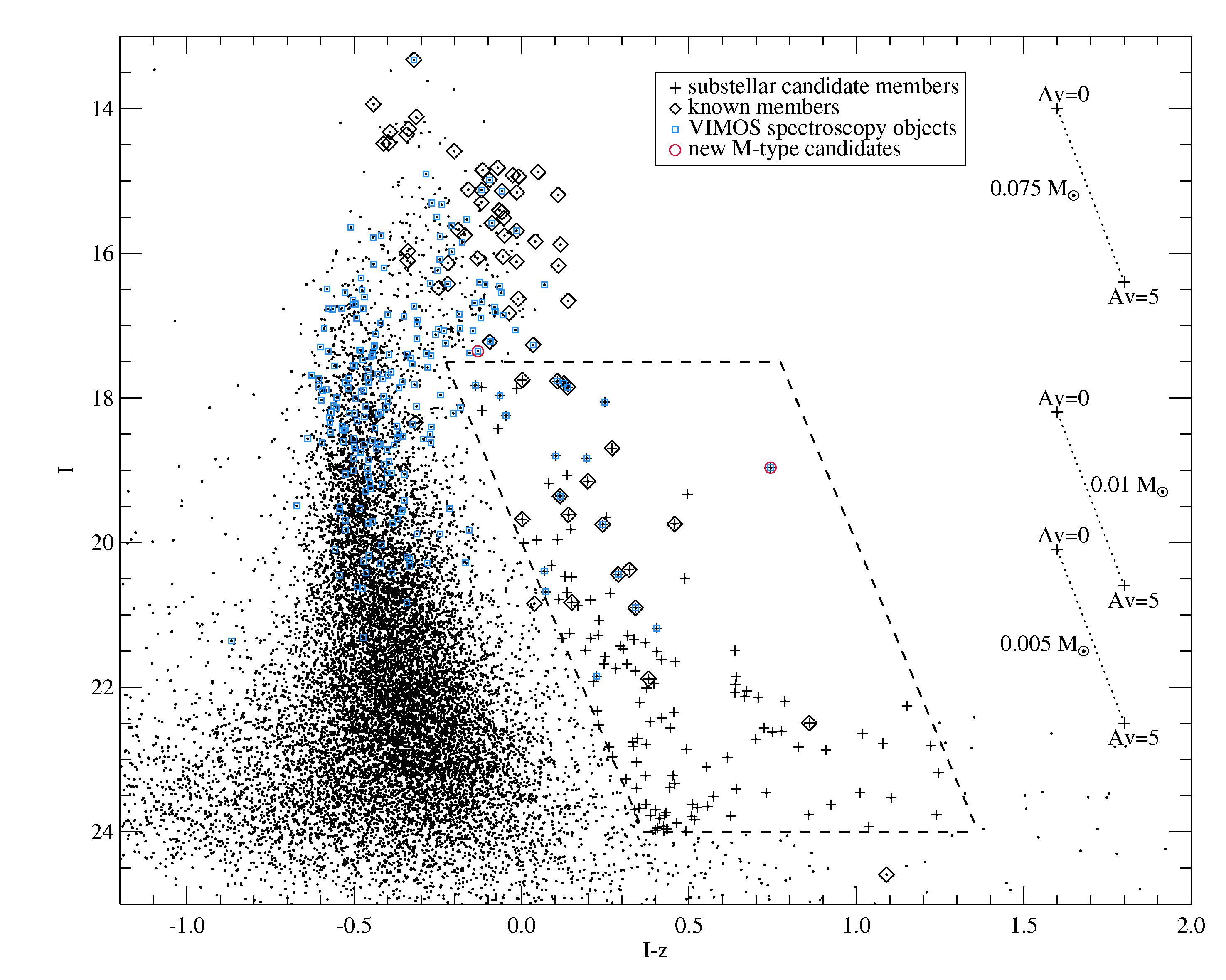}
\caption{$I$ vs $I-z$ color-magnitude diagram of the four Cha-I fields observed with VIMOS. The plus signs mark the substellar candidate sample, and small blue squares the objects with available optical spectra. Two M-dwarf candidates 
identified in this work are shown as red circles. Diamonds mark the confirmed M-type members objects according to the 
Luhman census. On the right-hand side of the diagram, we plot the expected $I$-band brightness of Cha-I members with the mass 0.075, 0.015, and 0.005 $M_{\odot}$, each at $A_V$=0 and 5; based on COND03 and DUSTY00 evolutionary models.}
\label{optCMD}
\end{figure*}

The (I, I-z) color-magnitude diagram of the sources observed in the four Cha-I fields is shown in Figure \ref{optCMD}. 
Previously spectroscopically confirmed M-type members (Luhman census) are shown as diamonds. On the right-hand side of the diagram, we plot the expected $I$-band brightness of Cha-I members with the mass 0.075, 0.015, and 0.005 $M_{\odot}$, each at $A_V$=0 and 5. For the calculation we use  COND03 \citep{baraffe03} and DUSTY00 \citep{chabrier00} evolutionary models. 

Very-low-mass (VLM) objects are expected to occupy a distinct region on the red side of the broad cumulation of the background main-sequence stars. We construct the selection box in the CMD such that it encompasses the majority of the known members of the spectral type M6 or later. 142 sources found within the selection box served as candidates for spectroscopic follow-up with VIMOS.

The faintest confirmed member in the color-magnitude diagram has $I=24.6$. We identify this source as 2M~J11070369-7724307. \citet{luhman08}
published the spectrum, and estimated spectral type M$7.5\pm1$ and effective temperature of 2795~K. The source suffers the extinction of $A_V\sim15-16$ (estimate from \citealt{luhman08} and our $J-K$ color, assuming the intrinsic $J-K=1$), which is higer than the average extinction in Cha-I (see Section \ref{missing}). According to DUSTY00 and COND03 models, a 2550~K ($\sim$M8.5) object at the distance of Cha-I and $A_V=16$ is expected to have $I=23.7-24.8$, in agreement with our photometry.

\subsection{Spectroscopic follow-up}
\label{s32}

\begin{figure}
\center
\includegraphics[width=7.5cm,angle=0]{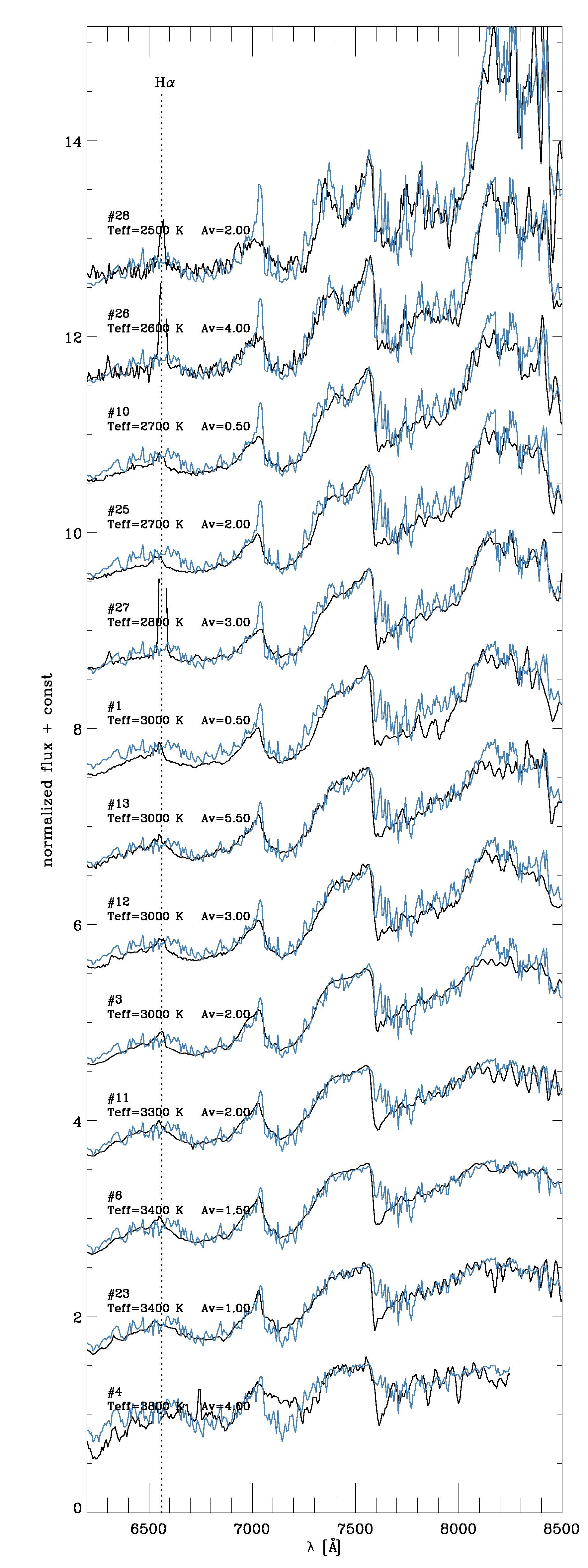}
\caption{Spectra of the objects in Table~\ref{tspec} (black), and the best-fit model (blue; AMES-Dusty; \citealt{allard01}). The spectral resolution of the models has been reduced 
by boxcar smoothing, in order to match the resolution of our spectra. Each spectrum has been corrected
for extinction, but not for the atmospheric absorption.}
\label{spectra}
\end{figure}

Slits were allocated to 60 out of 142 objects selected from the optical CMD, using the VIMOS Mask Preparation Software (VMMPS). A number of slits were also placed on the already known M dwarfs in the region. To fully use the multiplexing capability of the spectrograph, remaining slits were placed on the randomly chosen objects
within the FOV. 
As already mentioned above, some slits appear to be empty, or have poor signal-to-noise ratio.      
Blue squares in Figure $\,$\ref{optCMD} show 229 objects with the VIMOS spectra that can be used
for spectral classification; 18 of those
are found within the selection box.
The approximate cutoff for spectroscopy is evident from the CMD, as the spectra of the vast majority of 
the objects fainter than $I=21$ exhibit S/N too low to allow conclusive classification.
 
Our goal is to identify young members of Cha-I with masses in the substellar regime, which
typically corresponds to objects with effective temperature below 3000$\,$K. The photometrically chosen sample
might be contaminated by embedded stellar members of Cha-I, reddened background M-type stars, and
(less likely) late M- and early L-type objects in the foreground.
The red portion of M-dwarf spectra is dominated by molecular features (mainly TiO and VO; \citealt{kirkpatrick91,kirkpatrick95}), which allows a relatively simple preliminary selection of candidates based on visual inspection. Here we included all the objects with a clear evidence, 
but also, for the sake of caution, those objects with only a tentative evidence of molecular features. 
Another property that was used in the selection is the slope, 
which should be positive in the red portion of the optical regime for M-type stars. Earlier stellar types typically have negative slopes, or appear flat (late K-type). Among the rejected spectra we mainly find flat spectra with 
no, or with very weak indication of molecular features (late K-dwarfs, or reddened earlier-type stars), and
reddened featureless spectra (background or embedded stars of spectral type earlier than K). 

After this preliminary selection,
our sample contains 28 spectra.
To determine the effective temperature of our candidate objects we performed spectral fitting
using the AMES-Dusty models \citep{allard01}. Model spectra are previously smoothed by a boxcar function
to match the wavelength step of the VIMOS spectra.
We compared the effect of log$\,g$ in range 3.5--6.0 for constant effective temperature and found 
that gravity does not significantly change the shape of the optical low-resolution spectra. 
We search for the lowest $\chi^2$ by varying the effective temperature $T_{\mathrm{eff}}$,
and the interstellar extinction $A_V$, for the lowest available log$\,g\,$ (3.5 or 4.0).
We vary $T_{\mathrm{eff}}$ between 2000$\,$K and 3900$\,$K, in steps of 100$\,$K, and $A_V$
from 0 to 6, in steps of 0.5~mag.
To de-redden our spectra, 
we apply the extinction law from \citet{cardelli89}, assuming $R_V\,$=$\,$4. 
Due to the low resolution, it is the overall shape and the broad features in the spectrum 
that finally determine the best fit. To keep the fitting range
as broad as possible, we set the wavelength limits at 6200 and 8500~\AA. 
The cut at the red side is set to exclude the 
fringe-dominated portion of the spectra, while the cut at 6200~\AA~is set to avoid
the problematic region around 6100$\,$\AA, where the models typically overestimate the observed flux. 
The results of our spectral fitting are given in Table~\ref{tspec}, and shown 
in Figure~\ref{spectra}. We show only the objects with the best-fit
$T_{\mathrm{eff}}$ at or below 3800$\,$K 
because the objects found at the edge of the fitting grid could be even warmer.
The uncertainty is 100$\,$K for the
derived effective temperature, and 0.5$\,$mag 
for the extinction, 
which reflects the spacing of the grid. 

The models predict a much sharper peak at $\sim$7000~\AA~ than apparent in our spectra, a discrepancy
that becomes more pronounced for lower temperatures. 
According to \citet{kirkpatrick91}, the TiO bandhead at 7050~\AA~ is found in all late-K to late-M spectra, but is
strongest in early- to mid-M objects. \citet{reid95} suggest that the depth of this
feature provides a reliable spectral typing scheme for K7 -- M6.5 dwarfs. At later types, however, the depth
of the TiO feature decreases, which reflects both saturation in the TiO bandheads, and the presence of increasingly
strong VO bandheads \citep{reid95}. This behaviour is observed in our spectra, but is not well reproduced by the models.
In the same spectral region, object \#4 
shows flux levels above the model prediction and in this aspect stands out from the rest of the spectra in Figure$\,$\ref{spectra}. This might indicate that the object is hotter than our best-fit temperature at 3800~K.

We have to point out that the spectra shown in Figure~\ref{spectra} are not
corrected for telluric absorption. However, the two above mentioned spectral regions where models
do not fit the spectra well (6100 and 7050~\AA) are free from atmospheric bands. The only region where the effects
of telluric absorption are evident, is a deep O$_2$ absorption at $\sim7600\,$\AA.

The majority of the 
sources (11/13) shown in Figure~\ref{spectra} are previously identified members of Cha-I according to the Luhman census.
The effective temperatures obtained by our model fits are in agreement with those
from the literature. The mean of the difference between the two sets of $T_{\mathrm{eff}}$ is 20$\,$K, with
a standard deviation of 150$\,$K.
The two newly discovered objects with M-dwarf spectral features have 
effective temperatures of 3400 and 3800$\,$K, placing them among the
early M-type objects. The extinction estimate for these two sources lies in the range $A_V=1-3$ which is consistent
with the extinction towards Chamaeleon. However, without other clear signatures of youth, we cannot
yet claim cluster membership. Further spectroscopy is needed to clarify their status.  

\begin{deluxetable*}{cllccll}
\tabletypesize{\scriptsize}
\tablecaption{Parameters of Cha-I M-type objects observed with VIMOS \label{tspec}}
\tablewidth{0pt}
\tablehead{\colhead{ID} & \colhead{$\alpha$(J2000)} & \colhead{$\delta$(J2000)} &
\colhead{$T_{\mathrm{eff}}$ (K)} & \colhead{A$_{V}$ (mag)} & \colhead{Reference\tablenotemark{a}} & \colhead{name}}
\tablecolumns{7}
\startdata
  1   &  11 09 45.1833 & -77 40 33.4442 & 3000 & 0.5 & L07 & ESO$\,$H$\alpha\,$566\\
  3   &  11 08 16.9700 & -77 44 11.6052 & 3000 & 2.0  & L04 & Cha$\,$H$\alpha\,$13\\
  4   &  11 10 18.7603 & -77 46 13.8007 & 3800 & 4.0  & this work &\\
  6   &  11 10 11.5437 & -77 33 51.9489 & 3400 & 1.5 & L07\\
  10   & 11 12 02.7459 & -77 22 48.1824 & 2700 & 0.5 & L07 &\\
  11   & 11 10 50.7099 & -77 18 03.2794 & 3300 & 2.0 & L07 & ESO$\,$H$\alpha\,$568\\
  12   & 11 07 59.9412 & -77 15 31.9702 & 3000 & 3.0 & L07 & ESO$\,$H$\alpha\,$561\\
  13   & 11 08 55.9646 & -77 27 13.2001 & 3000 & 5.5 & L07 & ISO$\,$167\\
  23   & 11 13 01.2000 & -76 32 54.0509 & 3400 & 1.0 & this work &\\
  25   & 11 08 51.7863 & -76 32 50.4529 & 2700 & 2.0 & L04, L07 &\\
  26   & 11 08 49.5341 & -76 38 44.0479 & 2600 & 4.0 & L07 & \\
  27   & 11 09 52.1734 & -76 39 12.6947 & 2800 & 3.0 & L04&  ISO 217\\
  28   & 11 10 06.5998 & -76 42 48.4937 & 2500 & 2.0 & L07 &
\enddata
\tablecomments{Based on AMES-Dusty models \citep{allard01}. Spectra of the objects in this table, together with
the best-fit models are shown in Figure~\ref{spectra}.} 
\tablenotetext{a}{References for the published spectra: L04:~\citet{luhman04}, L07:~\citet{luhman07}}
\label{tspec}
\end{deluxetable*}

\subsection{Candidates from near-infrared and Spitzer photometry}
\label{s34}

\begin{figure}
\center
\includegraphics[width=8cm, angle=0]{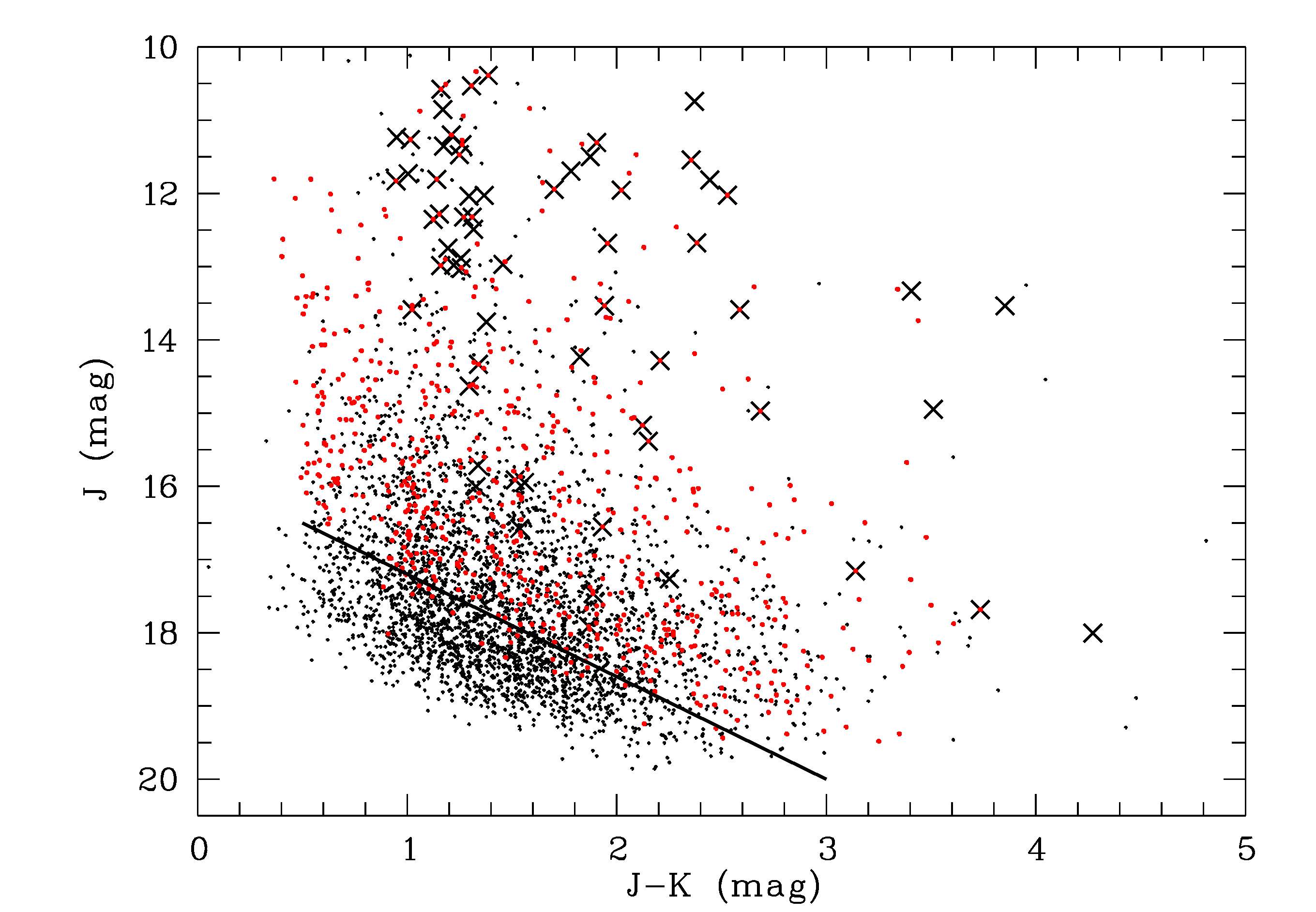}
\caption{Near-infrared color-magnitude diagram for the sources in Cha$\,$I.
Known members are marked with crosses; objects with Spitzer counterpart are shown in red. The solid
line shows the approximate limit of the combined $JK-Spitzer$ catalog.}
\label{NIRCMD}
\end{figure}

\begin{figure}
\center
\includegraphics[width=8cm, angle=0]{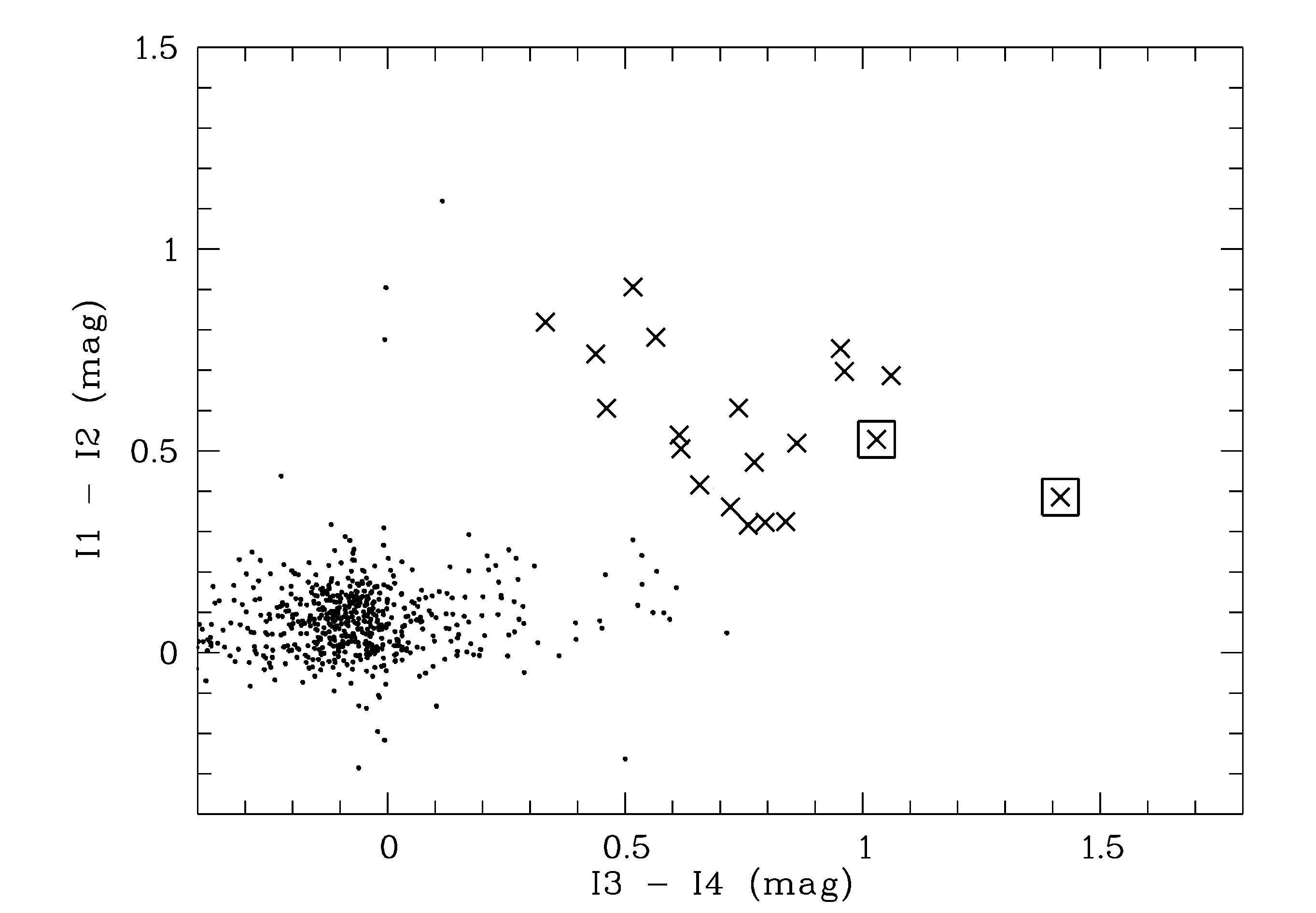}
\caption{IRAC color-magnitude diagram for the sources detected in $J$- and $K_S$-band. 
Crosses show 20 objects with color excess typical for objects with disks
and/or envelopes. Two of these sources do not yet have 
clarified membership (squares).}
\label{MIRCMD}
\end{figure}

\begin{deluxetable*}{cllccccccl}
\tabletypesize{\scriptsize}
\tablecaption{Candidate members in Cha-I from $JK/Spitzer$ photometry \label{tJKSp}}
\tablewidth{0pt}
\tablehead{
\colhead{ID} & \colhead{$\alpha$(J2000)} & \colhead{$\delta$(J2000)} &
\colhead{J (mag)}  & \colhead{J-K (mag)} &
\colhead {3.6\,\micron} & \colhead{4.5\,\micron} & \colhead{5.8\,\micron} & \colhead{8.0\,\micron} & \colhead{Notes} }
\tablecolumns{10}
\startdata
IR1   & 11 08 45.38 & -77 33 03.2    & 17.956 & 2.106  &    15.298 &  14.912 &  14.877 &  13.461 &  extended? \\
IR2 & 11 11 23.87 & -76 42 29.1    & 18.195 & 2.142  &    14.847 &  14.319 &  13.945 &  12.916 &  extended? 
\enddata
\end{deluxetable*}

The second candidate selection was carried out based on our near-infrared catalog, which was complemented by Spitzer/IRAC data, as outlined in Sect. \ref{s23}. In total, our catalog has 3768 sources detected in $J$- and $K_S$-band. 64 of them are known members of Cha-I, according to the Luhman census. 

In this catalog are 643 objects with Spitzer/IRAC counterpart within 3\farcs0, most of them (521) in the deep images from program \#36.  In Figure \ref{NIRCMD} we show the $(J, J-K)$ color-magnitude diagram generated from our catalog. Known Cha-I members are marked with crosses; objects with Spitzer counterpart are plotted in red. As can be seen in the diagram, the Spitzer survey misses many objects at the faint end of our $JK$ catalog. The approximate magnitude limit of the combined $Spitzer-JK$ catalog is shown with the solid line. The limit ranges from $J=17-20$ for $A_V = 0-10$\,mag. According to the COND03 and DUSTY00 isochrones, this limit corresponds to masses of $\sim 0.005\,M_{\odot}$.

To select new candidate members in Cha-I, we make use of the standard IRAC color-color diagram, as shown in Figure 
\ref{MIRCMD}. Plotted are the $5.8-8.0\,$\micron\ color vs. the $3.6-4.5\,\micron$ color for all 643 objects with Spitzer counterpart. As shown extensively in the literature, this diagram serves as a robust diagnostic for the presence of circumstellar dust around a particular object \citep{Allen04, Hartmann05}. Objects with disks and/or envelopes show excess in the two IRAC colors. The overwhelming majority of our sources appears around the origin of the plot, which precludes the presence of a disk. A small group of 20 objects has significant excess in both axes and is clearly distinct from the bulk of the datapoints (marked with crosses). 18 of them are known members according to the Luhman census. The remaining 2 are listed in Table \ref{tJKSp}, and marked with squares in Fig.~\ref{MIRCMD}. 
To our knowledge, these objects are new candidate members of Cha-I, although they both look slightly extended in the near-infrared images and could be galaxies. For a definite statement about the membership of the new candidates, spectroscopy is needed.

\section{Discussion}
\label{s4}

\subsection{Number of missing substellar objects}
\label{missing}
Based on our optical and $JK/Spitzer$ surveys we can put limits on the number of very low mass sources that are missing in the current census of YSOs in Cha-I.
In Figure \ref{spatial} we show the spatial coverage of our $IzJK$ survey (solid lines), the positions of the known members (black dots), including the known brown dwarfs (diamonds), the two candidates from Table \ref{tJKSp} (squares), and the two new candidate members from spectroscopy (circles). Our fields cover 88 out of 227 known members (38\%), from which 67 are recovered in our Iz catalog. 64 members are recovered in the $JK$ catalog, including 21 with disks.
Since the object density across Cha-I is highly variable, we consider it best to scale the numbers with the fraction of known objects covered by the survey, and not with the fraction of total area. The scaling factor for obtaining the number of potential new candidates in the regions not covered by our survey is then $1/(1-0.38)=1.6$.

\citet{cambresy99} derived extinction map of the Chamaeleon-Musca region based on optical star counts. Here we define
the extent of Cha-I as the region on the plane of the sky outlined by a circle of radius  $r=1.5^{\circ}$ and centered at ($\alpha\sim 11:05:00$, $\delta\sim -77:24:00$), and within the contours of $A_V\ge0.5$. Only $1\%$ of this area is at 
$A_V\ge5$. Therefore, when deriving mass equivalent to certain brightness limit, we consider 
only extinctions below $A_V=5$.

Let us consider the sources in the candidate selection box of the optical part of the survey in Figure$\,$\ref{optCMD}, that are brighter than the limit for spectroscopy, found at $I\approx\,21$. The upper limit placed at $I=17.5$ corresponds to masses of ($0.015-0.05$)$\,M_{\odot}$, and the lower limit $I=21$ to ($0.004-0.008$)$\,M_{\odot}$, in both cases for 
$A_V=0-5$. The estimates are based on DUSTY00 and COND03 evolutionary models.
From the 45 sources in the box, 15 are known members according to the Luhman census. We obtained spectra for 9 of the remaining 30 objects, and find 1 potential new M-type member, falling above the substellar temperature limit. From this we conclude that there are potentially 2 additional missing members in the $I=17.5-21$ range, in the surveyed part of the cluster. The total number of the missing VLM members with $I<21$ in Cha-I is thus $3\times 1.6 + 2 =7$ (unsurveyed + surveyed parts of the cluster).      

In our $JK/Spitzer$ survey,
we find zero new objects with disks brighter than $J=17$\,mag, and 2 potential new members with disks fainter than $J=17$\,mag. According to the COND03 and DUSTY00 models, the $J=17$ limit corresponds to (0.005-0.008)$\,M_{\odot}$ for $A_V=0-5$, similar to the $I=21$ limit. 
These numbers have to be scaled with the disk fraction ($\sim 50$\%, according to \citealt{luhman08}) and the coverage.   The total number of potentially missing very low mass members in Cha-I is $<3$ for $J<17$ and $<7$ for $J>17$\,mag.
This estimate relies on three assumptions:

a) The disk fraction does not change as a function of mass. 
In Cha-I, \citet{luhman08} find that the disk fraction 
is roughly constant at $\sim$50\% from 0.01 to 0.3$M_{\odot}$. 
This is not necessarily the case in other star forming regions, as shown, for example, by \citet{lada06} and \citet{scholz07}. Below the deuterium-burning limits, the fraction of the disk-bearing objects is even more uncertain.

b) The spatial distribution does not change as a function of mass. It is conceivable that brown dwarfs are scattered towards the outskirts of Cha-I whereas the higher mass stars are more concentrated towards the cloud core. From the current census, however, there is no evidence for brown dwarfs being distributed different than stars 
\citep{luhman07}.

c) The Spitzer survey fully covers our $JK$ area. In Figure \ref{spatial} we overplot the area on the deep and shallow IRAC images with dashed and dash-dotted lines, respectively. Our fields 1 and 4 with the highest density of known members (55/88) are completely covered by the deep IRAC images. The shallow IRAC images cover about 5/8 of the remaining fields. However, in these regions the density of known members is significantly lower and they are offset from the two main cloud cores in Cha-I. Therefore we do not consider the missing coverage a serious limitation.

In summary, from both the optical and the $JK/Spitzer$ survey we conclude that the current census of members in Cha-I is mostly complete down to $\sim0.008M_{\odot}$ for $A_V\leq5$. Within these limits and barring a significant mass dependence in disk fraction or spatial distribution, the number of missing object is $\leq\,7$. There may still be a significant number of unidentified members below this threshold or at higher extinctions.

\subsection{Ratio of stellar to substellar members}
\label{ratio}
Based on the current census in Cha-I, we can estimate the ratio of low-mass stars ($0.08M_{\odot}<m<1M_{\odot}$) 
to substellar objects. As substellar objects, we assume all objects with $T_{\mathrm{eff}}<3000\,$K, i.e. of the spectral
type M6 or later. There are 183 low-mass stars in the Luhman census. Here we do not include the
two new low-mass-star candidates found in this work, since their membership yet has to be confirmed.
The number of substellar objects in the current census is 40. 10 out of these 40 objects are of spectral type M9 or later 
($T_{\mathrm{eff}}\leq2400\,$K). 1$\,$Myr old objects at these low temperatures would have masses $\leq0.015M_{\odot}$, according to the DUSTY00 and COND03 models.
Ratio of low-mass-stars to BDs in Cha-I is therefore $183/30\approx6$, in agreement 
with the 3.3--8.5 limits found in other star forming regions (\citealt{andersen08}; calculated for the mass range at and above $0.03M_{\odot}$). 
This number represents the upper limit on the star-to-BD ratio, since we are probably still missing some objects. 

The ratio between the number of PMOs to BDs is currently $10/30=33\%$. This number, however, is subject to a 
large uncertainty, due to missing BDs and also a much larger number of missing PMOs.
From our statistics, we can give a crude estimate of the number of missing PMOs in Cha-I. The BD/PMO limit at $0.015M_{\odot}$ corresponds to $I\approx 17.1-19.5$ for $A_V=0-5$. For the objects with $I\leq19.5$ in the selection box in 
Figure~\ref{optCMD}, and following the same reasoning as in Section~\ref{missing}, we estimate that the number of missing objects is $\sim3$. This means that the total number of missing BDs is $\lesssim3$.
All remaining missing objects should be PMOs. Down to $I=21$ ($J=17$) 
this gives 4--7 PMOs. From our $JK/Spitzer$ survey in Section~\ref{missing} we estimated the number
of missing objects below this limit to be $<7$. In total, this estimate gives 4--13 missing PMOs in Cha-I.
The PMO-to-BD ratio in Cha-I then lies in the range between $42\%$ and $77\%$ while the ratio of PMOs to the total cluster population becomes $5-8\%$. 
We note that the ratios derived here should be taken with caution because they are subject to large uncertainties due to scaling from small numbers.

We can compare the derived numbers with those found in other star forming regions.
In $\sigma$~Ori, \citet{caballero07} find the lower limit for the PMO versus BD ratio of $25\pm^9_8\%$.   
In the ONC, \citet{lucas06} estimate the fraction of PMOs in the total cluster population to be
in range (1--14)$\%$. Highly variable extinction and age uncertainty make an unbiased census in the ONC difficult. 
While these numbers are in line with our Cha-I estimates, significantly different behavior is observed in NGC~1333. \citet{scholz09a} report an overabundance of BDs relative to low-mass stars (ratio of $1.5\pm0.3$), and a deficit
of PMOs. They find no PMOs in NGC~1333, although clearly being able to do so, as judged from the completeness limit of the survey. 
While the reason for this observed difference is not obviouos, it might indicate that the environment
may leave an imprint on the shape of the IMF at the lowest masses. The major contributions to the uncertainties
in the PMO/BD ratio are introduced by the completeness limits (most of the surveys are incomplete in the PMO regime), and the mass estimates that are based on the models, together with the age and distance uncertainties. High and variable 
extinction that is typical for most star forming regions additionally complicates this situation, as color-magnitude criteria used for candidate selection may also include more massive members at
higher extinctions. Here we must stress the importance of obtaining follow-up spectroscopy of the photometrically-selected candidates, because only the spectra can give us the information about the youth (and thus membership) and temperature of the observed objects.

As pointed out by \citet{luhman07}, and also apparent from Figure \ref{spatial}, the spatial distribution
of brown dwarfs does not significantly differ from the distribution of stars in Cha-I. This indicates that the 
ejection from multiple systems cannot be the dominant process of brown dwarf formation in this cluster. Any other
statement concerning the relative importance of BD formation scenarios is not possible based only on ratios
of BDs to stars or PMOs to BDs. To some extent, any of the proposed models can successfully reproduce 
the observed IMF. For example, in
simulations of the turbulent cloud fragmentation by \citet{padoan&nordlund04}, abundances of brown dwarfs in different
star forming regions can be matched by an appropriate combination of cloud density, temperature and the Mach number of the turbulence. To constrain the relative contributions of different formation mechanisms to the observed substellar population 
in different star forming regions, we need secondary indicators that provide a more effective ways to distinguish between 
different scenarios (see e.g. \citealt{bonnell07}). These secondary indicators include spatial distribution of stars and BDs in star forming regions, possible kinematic differences between the two populations, and binary properties (frequency, separations, mass ratios) and how they depend on the primary mass \citep{ahmic07, lafreniere08}.

\section{Summary and conclusions}
\label{s5}
This paper is the third in the series of papers presented in the framework of the SONYC project (short for 
Substellar Objects in Nearby Young Clusters). We have obtained deep optical and near-infrared observations of 4 fields in  the Chamaeleon-I star forming region, using VIMOS at the ESO VLT and SOFI at the ESO NTT. Follow-up spectroscopy of the candidates selected from the photometry was obtained using VIMOS/VLT. In the following, we summarize the survey and its most important results.

\begin{enumerate}
\item In our imaging survey, we reach completeness limits of 23.0 in $I$, 18.3 in $J$, and 16.7 in $K_S$. This corresponds 
to mass limits of $(0.003 - 0.005) M_{\odot}$, for $A_V=0-5$ and the distance of $\sim\,$160$\,$pc (based on 
DUSTY00 and COND03 models). This is more than 2$\,$magnitudes deeper in all three bands than the completeness limits of the major part of the currently most extensive Cha-I survey presented in \citet{luhman07}. Only one field in \citet{luhman07}, covering $0.22^{\circ} \times 0.28^{\circ}$ (equivalent to a quarter of our coverage), did achieve photometric depths comparable to those presented here.
\item The spatial coverage of our survey is 0.25$\,$deg$^2$. It includes 38\% of the previously known members of Cha-I from the Luhman census. 
\item From the optical photometry, we select 142 objects with $I-z$ colors as expected for substellar members of the cluster. 45 of these objects are above the $I=21$ limit observed in our spectroscopic follow-up, and for 18 of these sources we have obtained optical spectra. We have also obtained spectra for 211 randomly selected objects outside the selection box, including several known members. We find two objects with spectra consistent
with the spectral type M, but do not have enough information to clarify membership.
None of the two new potential members is substellar, as the temperature estimate resulting from the spectral fitting are well above 3000$\,$K. 
\item Our JK photometric survey combined with the Spitzer data yields 2 new candidate members. The membership for these objects has to yet be confirmed by spectroscopy.
\item Based on the results of our survey, while scaling for the coverage and disk fraction, we estimate that the number of the missing low-mass members down to $\sim 0.008\,M_{\odot}$ and $A_V\leq5$ in Cha-I is $\leq\,7$, i.e. $\leq3\%$ of the total number of members according to the current census. We might, however, still miss objects with lower masses, and objects at higher extinctions.
\end{enumerate}

\acknowledgments
The research was supported in part by grants from
the Natural Sciences and Engineering Research Council
(NSERC) of Canada to RJ.
RJ also acknowledges support from a Royal Netherlands Academy of Arts
and Sciences (KNAW) visiting professorship.
This work was also supported in part by the Science Foundation Ireland  
within the Research Frontiers Programme under grant no. 10/RFP/AST2780.
This publication makes use of data products from the Two Micron All
Sky Survey, which is a joint project of the University of Massachusetts
and the Infrared Processing and Analysis Center/California Institute of
Technology, funded by the National Aeronautics and Space Administration
and the National Science Foundation.


\begin{thebibliography}{43}
\expandafter\ifx\csname natexlab\endcsname\relax\def\natexlab#1{#1}\fi

\bibitem[{{Ahmic} {et~al.}(2007){Ahmic}, {Jayawardhana}, {Brandeker}, {Scholz},
  {van Kerkwijk}, {Delgado-Donate}, \& {Froebrich}}]{ahmic07}
{Ahmic}, M., {Jayawardhana}, R., {Brandeker}, A., {Scholz}, A., {van Kerkwijk},
  M.~H., {Delgado-Donate}, E., \& {Froebrich}, D. 2007, \apj, 671, 2074

\bibitem[{{Allard} {et~al.}(2001){Allard}, {Hauschildt}, {Alexander},
  {Tamanai}, \& {Schweitzer}}]{allard01}
{Allard}, F., {Hauschildt}, P.~H., {Alexander}, D.~R., {Tamanai}, A., \&
  {Schweitzer}, A. 2001, \apj, 556, 357

\bibitem[{{Allen} {et~al.}(2004){Allen}, {Calvet}, {D'Alessio}, {Merin},
  {Hartmann}, {Megeath}, {Gutermuth}, {Muzerolle}, {Pipher}, {Myers}, \&
  {Fazio}}]{Allen04}
{Allen}, L.~E., {Calvet}, N., {D'Alessio}, P., {Merin}, B., {Hartmann}, L.,
  {Megeath}, S.~T., {Gutermuth}, R.~A., {Muzerolle}, J., {Pipher}, J.~L.,
  {Myers}, P.~C., \& {Fazio}, G.~G. 2004, \apjs, 154, 363

\bibitem[{{Andersen} {et~al.}(2008){Andersen}, {Meyer}, {Greissl}, \&
  {Aversa}}]{andersen08}
{Andersen}, M., {Meyer}, M.~R., {Greissl}, J., \& {Aversa}, A. 2008, \apjl,
  683, L183

\bibitem[{{Baraffe} {et~al.}(2003){Baraffe}, {Chabrier}, {Barman}, {Allard}, \&
  {Hauschildt}}]{baraffe03}
{Baraffe}, I., {Chabrier}, G., {Barman}, T.~S., {Allard}, F., \& {Hauschildt},
  P.~H. 2003, \aap, 402, 701

\bibitem[{{Bate}(2009)}]{bate09a}
{Bate}, M.~R. 2009, \mnras, 392, 590

\bibitem[{{Bertin} \& {Arnouts}(1996)}]{SExtractor96}
{Bertin}, E. \& {Arnouts}, S. 1996, \aaps, 117, 393

\bibitem[{{Bonnell} {et~al.}(2007){Bonnell}, {Larson}, \&
  {Zinnecker}}]{bonnell07}
{Bonnell}, I.~A., {Larson}, R.~B., \& {Zinnecker}, H. 2007, Protostars and
  Planets V, 149

\bibitem[{{Caballero} {et~al.}(2007){Caballero}, {B{\'e}jar}, {Rebolo},
  {Eisl{\"o}ffel}, {Zapatero Osorio}, {Mundt}, {Barrado Y Navascu{\'e}s},
  {Bihain}, {Bailer-Jones}, {Forveille}, \& {Mart{\'{\i}}n}}]{caballero07}
{Caballero}, J.~A., {B{\'e}jar}, V.~J.~S., {Rebolo}, R., {Eisl{\"o}ffel}, J.,
  {Zapatero Osorio}, M.~R., {Mundt}, R., {Barrado Y Navascu{\'e}s}, D.,
  {Bihain}, G., {Bailer-Jones}, C.~A.~L., {Forveille}, T., \& {Mart{\'{\i}}n},
  E.~L. 2007, \aap, 470, 903

\bibitem[{{Cambr{\'e}sy}(1999)}]{cambresy99}
{Cambr{\'e}sy}, L. 1999, \aap, 345, 965

\bibitem[{{Cardelli} {et~al.}(1989){Cardelli}, {Clayton}, \&
  {Mathis}}]{cardelli89}
{Cardelli}, J.~A., {Clayton}, G.~C., \& {Mathis}, J.~S. 1989, \apj, 345, 245

\bibitem[{{Carpenter}(2001)}]{carpenter01}
{Carpenter}, J.~M. 2001, \aj, 121, 2851

\bibitem[{{Chabrier} {et~al.}(2000){Chabrier}, {Baraffe}, {Allard}, \&
  {Hauschildt}}]{chabrier00}
{Chabrier}, G., {Baraffe}, I., {Allard}, F., \& {Hauschildt}, P. 2000, \apj,
  542, 464

\bibitem[{{Comer{\'o}n} {et~al.}(2004){Comer{\'o}n}, {Reipurth}, {Henry}, \&
  {Fern{\'a}ndez}}]{comeron04}
{Comer{\'o}n}, F., {Reipurth}, B., {Henry}, A., \& {Fern{\'a}ndez}, M. 2004,
  \aap, 417, 583

\bibitem[{{Comer{\'o}n} {et~al.}(1999){Comer{\'o}n}, {Rieke}, \&
  {Neuh{\"a}user}}]{comeron99}
{Comer{\'o}n}, F., {Rieke}, G.~H., \& {Neuh{\"a}user}, R. 1999, \aap, 343, 477

\bibitem[{{Devillard}(1997)}]{devillard97}
{Devillard}, N. 1997, The Messenger, 87, 19

\bibitem[{{Geers} {et~al.}(2010){Geers}, {Scholz}, {Jayawardhana}, {Lee},
  {Lafr\'eniere}, \& {Tamura}}]{geers10}
{Geers}, V.~C., {Scholz}, A., {Jayawardhana}, R., {Lee}, E., {Lafr\'eniere},
  D., \& {Tamura}, M. 2010, submitted to ApJ

\bibitem[{{Haisch} {et~al.}(2010){Haisch}, {Barsony}, \& {Tinney}}]{haisch10}
{Haisch}, K.~E., {Barsony}, M., \& {Tinney}, C. 2010, \apjl, 719, L90

\bibitem[{{Hartmann} {et~al.}(2005){Hartmann}, {Megeath}, {Allen}, {Luhman},
  {Calvet}, {D'Alessio}, {Franco-Hernandez}, \& {Fazio}}]{Hartmann05}
{Hartmann}, L., {Megeath}, S.~T., {Allen}, L., {Luhman}, K., {Calvet}, N.,
  {D'Alessio}, P., {Franco-Hernandez}, R., \& {Fazio}, G. 2005, \apj, 629, 881

\bibitem[{{Kirkpatrick} {et~al.}(1991){Kirkpatrick}, {Henry}, \&
  {McCarthy}}]{kirkpatrick91}
{Kirkpatrick}, J.~D., {Henry}, T.~J., \& {McCarthy}, Jr., D.~W. 1991, \apjs,
  77, 417

\bibitem[{{Kirkpatrick} {et~al.}(1995){Kirkpatrick}, {Henry}, \&
  {Simons}}]{kirkpatrick95}
{Kirkpatrick}, J.~D., {Henry}, T.~J., \& {Simons}, D.~A. 1995, \aj, 109, 797

\bibitem[{{Lada} {et~al.}(2006){Lada}, {Muench}, {Luhman}, {Allen}, {Hartmann},
  {Megeath}, {Myers}, {Fazio}, {Wood}, {Muzerolle}, {Rieke}, {Siegler}, \&
  {Young}}]{lada06}
{Lada}, C.~J., {Muench}, A.~A., {Luhman}, K.~L., {Allen}, L., {Hartmann}, L.,
  {Megeath}, T., {Myers}, P., {Fazio}, G., {Wood}, K., {Muzerolle}, J.,
  {Rieke}, G., {Siegler}, N., \& {Young}, E. 2006, \aj, 131, 1574

\bibitem[{{Lafreni{\`e}re} {et~al.}(2008){Lafreni{\`e}re}, {Jayawardhana},
  {Brandeker}, {Ahmic}, \& {van Kerkwijk}}]{lafreniere08}
{Lafreni{\`e}re}, D., {Jayawardhana}, R., {Brandeker}, A., {Ahmic}, M., \& {van
  Kerkwijk}, M.~H. 2008, \apj, 683, 844

\bibitem[{{L{\'o}pez Mart{\'{\i}}} {et~al.}(2004){L{\'o}pez Mart{\'{\i}}},
  {Eisl{\"o}ffel}, {Scholz}, \& {Mundt}}]{lopezmarti04}
{L{\'o}pez Mart{\'{\i}}}, B., {Eisl{\"o}ffel}, J., {Scholz}, A., \& {Mundt}, R.
  2004, \aap, 416, 555

\bibitem[{{Lucas} {et~al.}(2005){Lucas}, {Roche}, \& {Tamura}}]{lucas05}
{Lucas}, P.~W., {Roche}, P.~F., \& {Tamura}, M. 2005, \mnras, 361, 211

\bibitem[{{Lucas} {et~al.}(2006){Lucas}, {Weights}, {Roche}, \&
  {Riddick}}]{lucas06}
{Lucas}, P.~W., {Weights}, D.~J., {Roche}, P.~F., \& {Riddick}, F.~C. 2006,
  \mnras, 373, L60

\bibitem[{{Luhman}(2007)}]{luhman07}
{Luhman}, K.~L. 2007, \apjs, 173, 104

\bibitem[{{Luhman} {et~al.}(2008){Luhman}, {Allen}, {Allen}, {Gutermuth},
  {Hartmann}, {Mamajek}, {Megeath}, {Myers}, \& {Fazio}}]{luhman08}
{Luhman}, K.~L., {Allen}, L.~E., {Allen}, P.~R., {Gutermuth}, R.~A.,
  {Hartmann}, L., {Mamajek}, E.~E., {Megeath}, S.~T., {Myers}, P.~C., \&
  {Fazio}, G.~G. 2008, \apj, 675, 1375

\bibitem[{{Luhman} \& {Muench}(2008)}]{luhman&muench08}
{Luhman}, K.~L. \& {Muench}, A.~A. 2008, \apj, 684, 654

\bibitem[{{Luhman} {et~al.}(2004){Luhman}, {Peterson}, \& {Megeath}}]{luhman04}
{Luhman}, K.~L., {Peterson}, D.~E., \& {Megeath}, S.~T. 2004, \apj, 617, 565

\bibitem[{{Luhman} {et~al.}(2006){Luhman}, {Wilson}, {Brandner}, {Skrutskie},
  {Nelson}, {Smith}, {Peterson}, {Cushing}, \& {Young}}]{luhman06}
{Luhman}, K.~L., {Wilson}, J.~C., {Brandner}, W., {Skrutskie}, M.~F., {Nelson},
  M.~J., {Smith}, J.~D., {Peterson}, D.~E., {Cushing}, M.~C., \& {Young}, E.
  2006, \apj, 649, 894

\bibitem[{{Neuh{\"a}user} \& {Comer{\'o}n}(1999)}]{neuhauser&comeron99}
{Neuh{\"a}user}, R. \& {Comer{\'o}n}, F. 1999, \aap, 350, 612

\bibitem[{{Oasa} {et~al.}(1999){Oasa}, {Tamura}, \& {Sugitani}}]{oasa99}
{Oasa}, Y., {Tamura}, M., \& {Sugitani}, K. 1999, \apj, 526, 336

\bibitem[{{Padoan} \& {Nordlund}(2004)}]{padoan&nordlund04}
{Padoan}, P. \& {Nordlund}, {\AA}. 2004, \apj, 617, 559

\bibitem[{{Payne} \& {Lodato}(2007)}]{payne&lodato07}
{Payne}, M.~J. \& {Lodato}, G. 2007, \mnras, 381, 1597

\bibitem[{{Reid} {et~al.}(1995){Reid}, {Hawley}, \& {Gizis}}]{reid95}
{Reid}, I.~N., {Hawley}, S.~L., \& {Gizis}, J.~E. 1995, \aj, 110, 1838

\bibitem[{{Scholz} \& {Eisl{\"o}ffel}(2004)}]{scholz04}
{Scholz}, A. \& {Eisl{\"o}ffel}, J. 2004, \aap, 419, 249

\bibitem[{{Scholz} \& {Eisl{\"o}ffel}(2005)}]{scholz05}
---. 2005, \aap, 429, 1007

\bibitem[{{Scholz} {et~al.}(2009){Scholz}, {Geers}, {Jayawardhana}, {Fissel},
  {Lee}, {Lafreniere}, \& {Tamura}}]{scholz09a}
{Scholz}, A., {Geers}, V., {Jayawardhana}, R., {Fissel}, L., {Lee}, E.,
  {Lafreniere}, D., \& {Tamura}, M. 2009, \apj, 702, 805

\bibitem[{{Scholz} {et~al.}(2007){Scholz}, {Jayawardhana}, {Wood}, {Meeus},
  {Stelzer}, {Walker}, \& {O'Sullivan}}]{scholz07}
{Scholz}, A., {Jayawardhana}, R., {Wood}, K., {Meeus}, G., {Stelzer}, B.,
  {Walker}, C., \& {O'Sullivan}, M. 2007, \apj, 660, 1517

\bibitem[{{Stamatellos} \& {Whitworth}(2009)}]{stamatellos09}
{Stamatellos}, D. \& {Whitworth}, A.~P. 2009, \mnras, 392, 413

\bibitem[{{Whitworth} \& {Goodwin}(2005)}]{whitworth05}
{Whitworth}, A.~P. \& {Goodwin}, S.~P. 2005, Astronomische Nachrichten, 326,
  899

\bibitem[{{Zapatero Osorio} {et~al.}(2000){Zapatero Osorio}, {B{\'e}jar},
  {Mart{\'{\i}}n}, {Rebolo}, {Barrado y Navascu{\'e}s}, {Bailer-Jones}, \&
  {Mundt}}]{zapateroosorio00}
{Zapatero Osorio}, M.~R., {B{\'e}jar}, V.~J.~S., {Mart{\'{\i}}n}, E.~L.,
  {Rebolo}, R., {Barrado y Navascu{\'e}s}, D., {Bailer-Jones}, C.~A.~L., \&
  {Mundt}, R. 2000, Science, 290, 103

\end{thebibliography}

\clearpage

\end{document}